\documentclass[journal,twoside,web]{ieeecolor2}

\usepackage{generic}
\usepackage{cite}

\usepackage{amsmath,amssymb,amsfonts}

\usepackage{algorithm}
\usepackage{algorithmic}

\usepackage{graphicx}
\usepackage{svg}
\usepackage{subcaption}
\usepackage[export]{adjustbox}

\usepackage{textcomp}
\usepackage{xcolor}
\usepackage{array}
\usepackage{multirow}
\usepackage{booktabs}
\usepackage{comment}
\usepackage{capt-of}
\usepackage{float}
\usepackage{placeins}
\usepackage{balance}

\usepackage{hyperref}
\hypersetup{hidelinks}

\usepackage{scalerel}
\usepackage{tikz}
\usetikzlibrary{svg.path}

\def\BibTeX{{\rm B\kern-.05em{\sc i\kern-.025em b}\kern-.08em
    T\kern-.1667em\lower.7ex\hbox{E}\kern-.125emX}}
\providecolor{orcidlogocol}{RGB}{166,206,57}
\tikzset{
  orcidlogo/.pic={
    \fill[orcidlogocol] svg{M256,128c0,70.7-57.3,128-128,128C57.3,256,0,198.7,0,128C0,57.3,57.3,0,128,0C198.7,0,256,57.3,256,128z};
    \fill[white] svg{M86.3,186.2H70.9V79.1h15.4v48.4V186.2z}
                 svg{M108.9,79.1h41.6c39.6,0,57,28.3,57,53.6c0,27.5-21.5,53.6-56.8,53.6h-41.8V79.1z M124.3,172.4h24.5c34.9,0,42.9-26.5,42.9-39.7c0-21.5-13.7-39.7-43.7-39.7h-23.7V172.4z}
                 svg{M88.7,56.8c0,5.5-4.5,10.1-10.1,10.1c-5.6,0-10.1-4.6-10.1-10.1c0-5.6,4.5-10.1,10.1-10.1C84.2,46.7,88.7,51.3,88.7,56.8z};
  }
}
\newcommand\orcidicon[1]{\href{https://orcid.org/#1}{\mbox{\scalerel*{
\begin{tikzpicture}[yscale=-1,transform shape]
\pic{orcidlogo};
\end{tikzpicture}
}{|}}}}

\begin{document}
\title{Analysis of Autonomic Regulation in Cancer Survivors During Daily Physical Activity: A Real-World Wearable ECG Study}

\author{Sajad Farrokhi\orcidicon{0009-0009-8412-6930}, Lerick Sequeira\orcidicon{0009-0006-4371-0514}
, Shanna L. Burke\orcidicon{0000-0001-6969-3536}
, Waltenegus Dargie\orcidicon{0000-0002-7911-8081}, \IEEEmembership{Senior Member, IEEE}, and Christian Poellabauer\orcidicon{0000-0002-0599-7941}, \IEEEmembership{Senior Member, IEEE} 
    \thanks{Manuscript submitted on 15 June 2026.}
    \thanks{This work was partially supported by the Florida Department of Health, Florida Cancer Innovation Fund, under grants MOAAR and 25C38. The opinions, results, and conclusions are those of the authors and do not necessarily represent the official views of the Florida Department of Health.}
	\thanks{S. Farrokhi and C. Poellabauer are with  the Knight Foundation School of Computing and Information Sciences at Florida International University, USA, (e-mail: sfarrokh@fiu.edu, cpoellab@fiu.edu)}
    \thanks{L. Sequeira and S. Burke are with the School of Social Work at Florida International University, USA, (e-mail: lsequeir@fiu.edu, sburke@fiu.edu}
   \thanks{W. Dargie is with the Faculty of Computer Science, Technische Universit{\"a}t Dresden, 01062 Dresden, Germany (e-mail: waltenegus.dargie@tu-dresden.de)}
}

\maketitle

\begin{abstract}
This study investigates heart rate (HR) and heart rate variability (HRV) responses to physical activity in breast cancer survivors using wearable electrocardiogram (ECG) data collected in real-world settings. Reliable HRV analysis in such environments is challenging due to motion artifacts and activity-related signal degradation. To address this, we use an approach that combines accelerometer and gyroscope data for activity intensity segmentation (light, moderate, vigorous) with a robust ECG processing pipeline incorporating R-peak detection and annotation-free signal quality assessment. Because vigorous activity produced unreliable HRV estimates, analyses focused on light and moderate activity levels. Using 30~s, 1~min, and 2~min windows, HR and HRV metrics were computed and compared between breast cancer survivors and healthy controls. Cancer survivors consistently exhibited elevated HR and reduced HRV across activity levels. During light activity, HR increased from 95.7~bpm in controls to 103.4~bpm in cancer survivors. Differences became more pronounced during moderate activity, where RMSSD decreased from 39.7~ms to 22.1~ms and SDNN from 42.6~ms to 25.1~ms. Statistical analyses showed significant group differences with strong and consistent effects across observations. In addition, the proposed ECG quality assessment framework reliably identified high-quality signal segments, achieving near-perfect valid RR ratios (0.99) without manual annotations. Overall, these findings demonstrate impaired and activity-dependent autonomic regulation in cancer survivors and highlight the importance of motion-aware activity segmentation and robust ECG quality control for accurate physiological monitoring in real-world wearable settings.
\end{abstract}


\section{Introduction}

Heart rate (HR) and heart rate variability (HRV) are well-established indicators of autonomic nervous system function, providing critical insights into physiological responses under varying conditions~\cite{thayer2012meta}. They are of particular importance for the study of populations whose autonomic regulation may be impaired by specific diseases and medical treatments~\cite{zhou2016heart}. Breast Cancer survivors represent one such population.

Wearable electrocardiogram (ECG) technologies enable continuous monitoring of cardiovascular activity in real-world environments~\cite{zhang2025opportunities}. By moving beyond controlled clinical settings, real-world data captures natural variations in activity and stress, offering a richer and more realistic representation of daily physiological behavior. Nevertheless, ECG signals collected in real-world conditions are highly susceptible to motion artifacts and activity-induced variability, which can significantly degrade signal quality. These distortions hinder reliable detection of cardiac events and introduce errors in derived HR and HRV metrics~\cite{clifford2012signal, satija2018review}. The problem becomes pronounced when short ECG segments are used for analysis~\cite{raimondi2025effects}. This issue is further exacerbated during periods of movement or rapidly changing exertion levels. Furthermore, the lack of ground truth annotations in large-scale wearable datasets makes it challenging to assess signal reliability and validate extracted features~\cite{orphanidou2017signal}. Consequently, robust and scalable HRV analysis in real-world environments remains an ongoing challenge.

\begin{figure}[!htbp]
\centering
\includegraphics[width=0.7\linewidth]{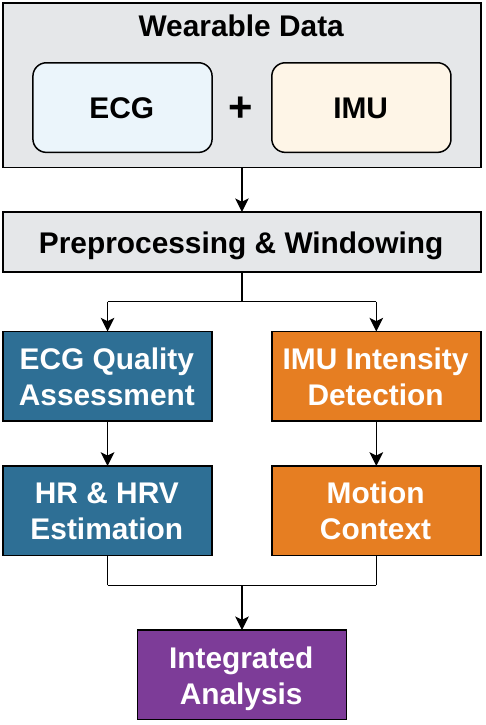}
\caption{Overview of the proposed quality-aware multimodal approach for HR and HRV analysis in real-world environments.}
\label{fig:framework}
\end{figure}

Knowledge of motion context, signal reliability, and exertion levels is helpful to improve signal selection, analysis, and interpretation. For example, measurements from inertial measurement units (IMUs) can provide reliable estimates of motion intensity~\cite{marin2020using}. Aligning these measurements with the measurements of  wearable electrocardiograms enables the identification of distortions in ECG segments and to reason about underlying causes. However, the state-of-the-art in ECG signal analysis and modeling offers limited insights when it comes to the joint investigation of physiological exertion, ECG signal quality, and heart rate variability~\cite{gaur2024continuous}. To address these limitations, we propose a unified, quality-aware, and multimodal approach for wearable ECG analysis with the main focus on autonomic dysfunction in breast cancer survivors. An overview of the proposed approach is illustrated in Figure~\ref{fig:framework}.

Specifically, we: 1) develop a unified, quality-aware framework that integrates ECG and inertial sensing for activity-aware HR and HRV analysis in real-world environments; 
2) propose an unsupervised clustering approach for activity intensity segmentation from accelerometer and gyroscope signals; 
3) introduce a clustering-based method for ECG signal quality assessment using signal features; 
4) design an annotation-free validation strategy based on physiologically grounded RR interval consistency to identify reliable segments; and 
5) evaluate the framework on a real-world dataset of breast cancer survivors and healthy controls, demonstrating that autonomic differences are strongly activity-dependent and most pronounced under moderate physical demand.

The remainder of this paper is organized as follows. Section~II reviews related work. Section~III presents the proposed methodology, including signal preprocessing, ECG quality assessment, activity intensity segmentation, and statistical analysis. Section~IV presents the results and discussion. Finally, Section~V concludes the paper.

\section{Related Work}
\subsection{ECG Signal Degradation and Motion Artifacts in Wearable Settings}

Wearable ECG monitoring systems are highly susceptible to signal degradation caused by motion artifacts, environmental interference, and acquisition-related variability. Raimondi et al.~\cite{raimondi2025effects} investigated the effect of additive noise on ultra-short-term HRV analysis by injecting controlled noise at different signal-to-noise ratio (SNR) levels, showing that HRV metrics become increasingly unreliable at SNR $\leq 5$~dB, with stable estimates observed only above 10~dB. Complementing this, Cui et al.~\cite{cui2024exploring} analyzed capacitive ECG acquisition by modeling the skin--electrode interface and experimentally evaluating factors such as environmental humidity and clothing thickness. Their results showed that increased skin-electrode impedance attenuates ECG amplitude and reduces waveform reliability, leading to degraded visibility of clinically important ECG features such as QRS complexes.

Similar challenges have been observed in real-world wearable ECG recordings, where motion artifacts caused by body movement, electrode displacement, and muscle activity, along with missing segments and device-dependent sampling inconsistencies, can distort ECG morphology, compromise R-peak detection, and affect downstream HRV analysis~\cite{RIMOK2026103377}. To mitigate these effects, recent studies have explored advanced denoising techniques, including deep learning methods such as generative adversarial networks (GANs) and CycleGAN frameworks. Rajasekhar et al.~\cite{rajasekhar2025dl} proposed a CycleGAN-based ECG denoising framework that improved ECG reconstruction quality under noisy conditions while preserving waveform morphology, achieving approximately 2--3~dB higher peak signal-to-noise ratio (PSNR) and lower mean squared error (MSE) compared to conventional GAN approaches. However, these methods are typically evaluated using controlled or simulated noise conditions, and their robustness under highly variable free-living motion artifacts remains less explored.

\subsection{ECG Signal Quality Assessment for Reliable HRV Analysis}

To improve the reliability of wearable ECG analysis, previous studies have explored rule-based, learning-based, and unsupervised approaches for ECG signal quality assessment. Clifford et al.~\cite{clifford2012signal} introduced signal quality indices (SQIs) based on agreement between multiple QRS detectors and RR interval consistency measures. Although effective in controlled datasets, the approach depends heavily on accurate peak detection and may degrade under severe motion artifacts. More recently, Zhang et al.~\cite{zhang2022deep} proposed a framework combining residual and recurrent neural networks for classifying ECG segments into good-, medium-, and poor-quality categories, achieving 92.31\% accuracy for three-class classification and 98.72\% for binary acceptable/unacceptable classification on the CPSC 2020 database containing 87,888 ECG segments, although reduced generalization was observed across heterogeneous populations. Similarly, Yu~\cite{yu2025research} extracted time- and frequency-domain ECG features and applied an unsupervised K-means clustering framework to separate signal-dominated and noise-dominated segments, improving signal-to-noise ratio up to 21.5~dB while preserving waveform similarity up to 95.8\%.

Beyond quantitative noise measures, Holgado-Cuadrado et al.~\cite{holgado2023characterization} proposed a clinically motivated ECG noise taxonomy based on waveform interpretability and achieved F1-scores of approximately 0.77 and precision up to 0.80 for distinguishing usable from non-usable ECG segments. Their findings demonstrated that segments with acceptable quantitative noise levels may still lack diagnostically meaningful waveform characteristics. In parallel, previous studies showed that even small inaccuracies in R-peak detection and RR interval estimation can substantially affect downstream HRV metrics, particularly in short-duration wearable recordings~\cite{peltola2012rr,berntson1997heart}. Collectively, these studies demonstrate that while existing methods can assess ECG signal quality effectively, most approaches do not explicitly evaluate the reliability of downstream HR and HRV analysis in free-living wearable environments.

\subsection{HRV, Physical Activity, and Cancer Survivorship}

Heart rate variability is strongly influenced by physical activity intensity and autonomic nervous system dynamics. To investigate these effects, Wang et al.~\cite{10.3389/fphys.2024.1462082} analyzed HRV responses during progressively increasing exercise intensity using a Bruce treadmill protocol in approximately 23 participants and reported substantial reductions in time-domain and frequency-domain HRV metrics with increasing physiological load. In particular, SDNN decreased from 40.43~ms to 8.70~ms, RMSSD decreased from 39.05~ms to 6.19~ms, and frequency-domain components such as LF and HF decreased by approximately 78\%, indicating a shift toward sympathetic nervous system dominance.

In clinical populations, HRV has been widely used to assess autonomic dysfunction and cardiovascular regulation. Several studies have shown that breast cancer survivors exhibit reduced HRV and elevated resting heart rate compared to healthy controls. For example, Caro et al.~\cite{caro2016heart} reported reductions in SDNN from 62.35 $\pm$ 18.21~ms in controls to 39.10 $\pm$ 16.28~ms in survivors and RMSSD from 58.65 $\pm$ 22.17~ms to 28.82 $\pm$ 23.52~ms. Similarly, Sousa et al.~\cite{sousa2024physical} observed reduced RMSSD values from 39.15 $\pm$ 37.66~ms in controls to 14.89 $\pm$ 8.28~ms in breast cancer patients, along with reductions in SDNN from 55.77 $\pm$ 40.03~ms to 26.30 $\pm$ 10.37~ms. Complementing these findings, Majerova et al.~\cite{majerova2022increased} analyzed time-domain, frequency-domain, and nonlinear HRV metrics in active cancer patients and survivors, reporting increased sympathetic modulation with 0V\% increasing from 16.17\% to 26.23\% ($p < 0.01$) and reduced signal complexity with sample entropy decreasing from approximately 1.86 to 1.60--1.62 ($p < 0.01$), although RMSSD and SDNN showed limited differences under controlled resting conditions. More broadly, Borsati et al.~\cite{borsati2025cardiac} reported that cardiac autonomic dysfunction may affect up to 80\% of cancer patients and survivors and associated these alterations with increased sympathetic activation, reduced vagal tone, systemic inflammation, and treatment-related cardiotoxicity.

Despite these findings, most HRV studies in cancer populations have been conducted under controlled resting conditions using short-duration recordings, without accounting for physical activity, posture, or real-world behavioral variability. Similarly, activity-aware HRV studies are typically performed under controlled laboratory conditions where signal quality is assumed to be reliable and motion-induced artifacts are minimized. As a result, limited work has jointly considered physical activity intensity, wearable ECG signal quality, and HRV estimation in real-world environments. Taken together, these limitations highlight the need for integrated frameworks that jointly consider motion context, ECG signal quality, and reliable HR/HRV estimation under realistic wearable sensing conditions.

\section{Methodology}
\subsection{Dataset}

The dataset consists of synchronized physiological and motion recordings collected using wearable devices in free-living conditions. Participants were recruited through targeted social media campaigns, and data collection was conducted under an approved Institutional Review Board (IRB) protocol at Florida International University (FIU) over a six-month period in 2025. All participants provided written informed consent prior to participation. The study included female breast cancer survivors aged 18 to 59 who had completed primary cancer treatment within the previous five years and resided in South Florida. Healthy controls met the same age criteria but had no history of cancer. Exclusion criteria for all participants included: (1) conditions impairing motor or cognitive function, (2) inability to independently operate a smartphone application, and (3) inability or unwillingness to use a wearable device. Healthy controls were additionally excluded if they had medical conditions with symptoms similar to dysautonomia, such as endocrine disorders, chronic fatigue syndrome, fibromyalgia, autoimmune disorders, or Ehlers-Danlos syndrome.

Following an initial in-person home visit, participants were trained on device setup, sensor placement, and recording procedures. The sensor was positioned at the center of the chest, with electrode pads moistened prior to placement to ensure optimal signal quality. Participants then conducted recording sessions in free-living environments of their choice, including homes, gyms, and other locations. Each participant completed at least three sessions per week, with a minimum duration of 15 minutes, covering a range of daily activities. In addition to free-living recordings, participants completed a supervised 1-minute seated resting baseline recording during the initial visit to provide a reference measure of resting cardiovascular activity.

Data were acquired using the Movesense MD wearable sensor (Model OP174; Movesense Oy, Espoo, Finland), a Bluetooth Low Energy-enabled Class IIa medical device certified under the European Union Medical Device Regulation (MDR 2017/745) \cite{movesense_md}. The sensor records single-lead ECG along with multi-axis IMU and magnetometer signals, enabling joint analysis of cardiac activity and motion. ECG signals were sampled at 125~Hz, while IMU and magnetometer data were sampled at 52~Hz. Participants connected the sensor to their smartphones, and recording sessions were initiated and terminated through the Movesense mobile application. In total, data were collected from 54 participants, including 37 breast cancer survivors and 17 healthy controls. The breast cancer survivor group had a mean age of 46.2$\pm$7.1 years, mean height of 163.3$\pm$5.8 cm, and mean weight of 71.9$\pm$15.3 kg, while the healthy control group had a mean age of 26.2$\pm$5.2 years, mean height of 163.1$\pm$6.9 cm, and mean weight of 63.5$\pm$12.7 kg.

Only segments with concurrent ECG and IMU recordings were retained, resulting in 292.55~hours of data. Due to variations in recording duration and participant compliance, the dataset is imbalanced, with a larger proportion of data from breast cancer survivors. To ensure reliability, the data were further refined through signal quality control and physiological validity constraints based on RR interval availability and HRV plausibility. After filtering, 142.49~hours of data remained for analysis, comprising 130.91~hours from the breast cancer survivor group and 11.58~hours from the control group. Table~\ref{tab:dataset_summary} summarizes the dataset size and data retention across processing stages.

\begin{table}[t]
\centering
\caption{Dataset size and data retention across processing stages.}
\label{tab:dataset_summary}
\setlength{\tabcolsep}{3pt}
\small
\begin{tabular}{lcccccc}
\toprule
\textbf{Group} & \textbf{N} & \textbf{Total (h)} & \textbf{QC (h)} & \textbf{Final (h)} & \textbf{QC (\%)} & \textbf{Final (\%)} \\
\midrule
Survivor & 37 & 272.46 & 225.71 & 130.91 & 82.84 & 48.05 \\
Control & 17 & 20.09  & 16.97  & 11.58  & 84.50 & 57.62 \\
\midrule
Total   & 54 & 292.55 & 242.68 & 142.49 & 82.95 & 48.71 \\
\bottomrule
\end{tabular}
\end{table}

\subsection{Signal Preprocessing}

The ECG signal was preprocessed using zero-phase Butterworth bandpass filtering, implemented via forward--backward filtering to avoid phase distortion~\cite{gustafsson2002determining}. The filter range was adapted to the analysis task. For signal characterization, clustering, and feature extraction, a 1--40~Hz band was applied to retain physiologically relevant ECG components while suppressing baseline drift and high-frequency noise, consistent with standard ECG preprocessing practices \cite{peltola2012rr}. In contrast, for R-peak detection, which forms the basis for subsequent HRV computation, a wider band of 1--60~Hz was used to preserve higher-frequency components of the QRS complex and improve peak localization accuracy under motion artifacts.

For IMU signals, accelerometer measurements were scaled when necessary to ensure consistent physical units. The gravitational component was removed using high-pass filtering with a cutoff frequency of 0.25~Hz, attenuating low-frequency components associated with device orientation and isolating dynamic acceleration due to motion, as commonly performed in wearable sensing applications \cite{bao2004activity}.

For clustering, extracted features were standardized to zero mean and unit variance:
\begin{equation}
x_{\mathrm{norm}} = \frac{x - \mu}{\sigma},
\end{equation}
where \( \mu \) and \( \sigma \) denote the mean and standard deviation of each feature.

Following preprocessing, both ECG and IMU signals were segmented into overlapping windows of 10~s with a 5~s step, enabling short-term analysis of physiological and motion dynamics while maintaining sufficient temporal resolution for capturing activity transitions. Additional task-specific preprocessing steps are described within the corresponding subsections.

\subsection{ECG Signal Quality Validation Without Annotations}

In real-world environments, ECG signals are frequently affected by motion artifacts and signal distortions, making reliable identification of physiologically meaningful segments challenging in the absence of manual annotations~\cite{ZHANG2025117067}. To address this, we adopt a data-driven approach in which ECG signal quality is first inferred through unsupervised clustering based on extracted ECG signal features. These features, summarized in Table~\ref{tab:ecg_features}, capture amplitude, spectral, and morphological characteristics commonly used in ECG signal analysis \cite{singh2023ecg}. The resulting clusters are subsequently validated using physiologically grounded metrics derived from R-peak detection.
\begin{table}[t]
\centering
%\scriptsize
\caption{ECG features used for signal quality characterization.}
\renewcommand{\arraystretch}{1.2}
\begin{tabular}{p{2cm} p{4cm}}
\hline
\textbf{Category} & \textbf{Features} \\
\hline
Amplitude & BP STD (1--40 Hz), RMS \\
Spectral & Low-band ratio, HF ratio \\
Morphology & Derivative RMS \\
\hline
\end{tabular}

\vspace{1mm}
\footnotesize{\textit{Abbreviations:} BP = bandpass; STD = standard deviation; RMS = root mean square; HF = high frequency.}

\label{tab:ecg_features}
\end{table}

Specifically, ECG signals were segmented into overlapping 10 s windows (5 s step) and grouped using unsupervised clustering based on signal characteristics. The resulting clusters are hypothesized to represent latent ECG signal quality levels. To evaluate clustering performance, standard validation metrics were computed, including the silhouette score~\cite{rousseeuw1987silhouettes} and Davies--Bouldin index~\cite{davies2009cluster}. The clustering achieved a silhouette score of 0.56 and a Davies--Bouldin index of 0.60, indicating well-defined and compact cluster structure, thereby supporting the validity of the derived ECG quality groups. However, because clustering is performed without supervision, an additional validation step is required to ensure that these groupings correspond to meaningful differences in physiological signal reliability. To this end, R-peaks were detected within each ECG window using the method described in Section~\ref{sec:feature_extraction}, and the detected peaks were used to derive a set of annotation-free metrics that assess the physiological plausibility and consistency of the resulting RR intervals.

The reliability of RR intervals was first evaluated based on their physiological range. The \textit{valid RR ratio} was defined as the proportion of intervals within 300--2000ms, consistent with established HRV preprocessing practices \cite{peltola2012rr, electrophysiology1996heart}, and serving as a primary indicator of reliable peak detection. Temporal consistency was further assessed using the \textit{RR instability ratio}, defined as the proportion of consecutive RR interval ratios falling outside predefined bounds (0.60--1.67). This metric evaluates how much the heart rate is allowed to change from one beat to the next. A ratio of 1 indicates no change, while values below 0.60 correspond to a sudden shortening of the RR interval by more than 40\%, and values above 1.67 correspond to a sudden lengthening by more than 67\%. Such large beat-to-beat changes are unlikely under normal physiological conditions, where heart rate typically varies gradually. Therefore, ratios outside this range indicate abrupt and non-physiological fluctuations, which are commonly associated with missed or spurious peak detections~\cite{berntson1997heart}. To further characterize signal stability, the \textit{coefficient of variation of RR intervals (RR CV)} was computed for each window, where higher values indicate increased variability and reduced signal reliability. In addition to temporal and statistical measures of RR variability, morphological consistency was subsequently assessed using \textit{beat-template correlation}, where individual beats were aligned and compared to a local template, and the average correlation was used as a measure of waveform consistency \cite{orphanidou2017signal}.

These metrics, summarized in Table~\ref{tab:ecg_quality_metrics} were aggregated within each cluster to evaluate whether clustering effectively separates ECG segments according to physiological signal reliability. Clusters exhibiting high valid RR ratios, low variability, and high morphological consistency were interpreted as \textit{high-quality} ECG segments suitable for HRV analysis. Clusters with intermediate metric values were interpreted as \textit{moderate-quality} segments, while clusters with degraded metric profiles were interpreted as \textit{low-quality} segments.

\begin{figure}[t]
    \centering
    \includegraphics[width=0.95\linewidth]{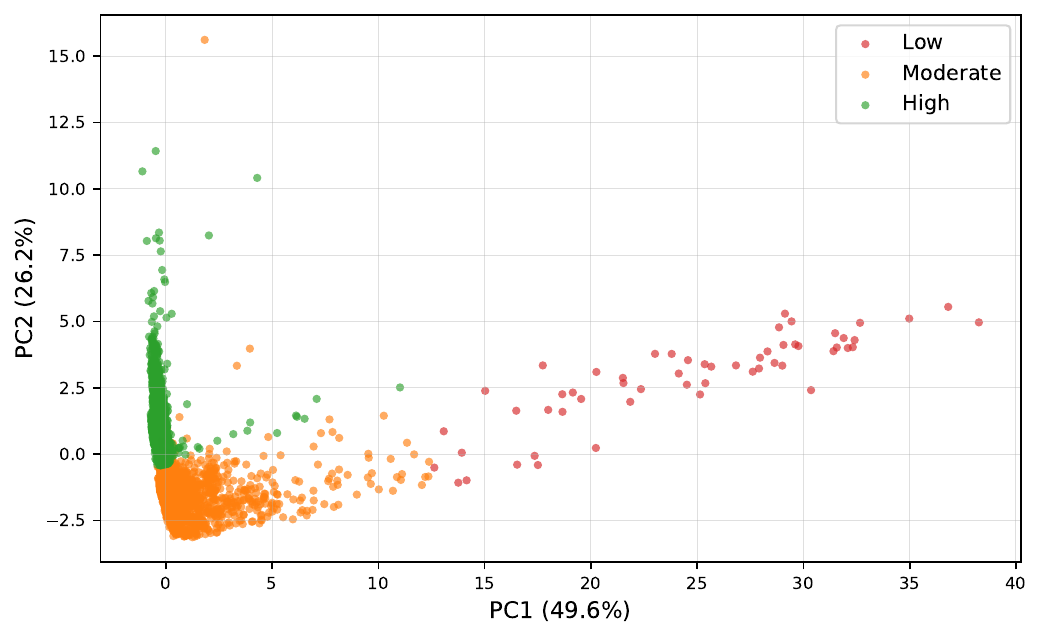}
    \caption{Projection of ECG windows in the first two principal components (PC1 and PC2) colored by cluster assignment. The clusters exhibit clear separation, indicating that the unsupervised grouping captures distinct signal regimes. The cluster corresponding to high-quality ECG segments forms a compact and well-structured region, whereas lower-quality clusters show greater dispersion and deviation along PC1, reflecting increased signal variability and distortion.}
    \label{fig:ecg_pca_clusters}
\end{figure}

It is important to note that these labels do not directly correspond to noise levels, but rather reflect the reliability and physiological plausibility of the detected cardiac activity within each cluster. As illustrated in Figure~\ref{fig:ecg_pca_clusters}, the clustered ECG windows exhibit clear separation in the feature space, supporting the assumption that unsupervised clustering captures distinct signal quality regimes. 

As an additional validation step, each 10~s ECG window was independently assigned a reliability label (high, moderate, or low) based on thresholding of the proposed quality metrics. Specifically, thresholds were applied to enforce physiological plausibility (valid RR ratio $\geq$ 0.90), low RR instability ratio $\leq$ 0.15, low RR variability (RR CV $\leq$ 0.10), and stable waveform morphology (beat-template correlation $\geq$ 0.70). Windows satisfying the majority of these criteria were labeled as high reliability, those satisfying a subset were labeled as moderate, and those failing most criteria were labeled as low reliability. 

The distribution of these independently assigned reliability labels across clusters is reported in Table~\ref{tab:ecg_quality_labels}, showing strong agreement for the high-quality cluster, which is predominantly composed of high-reliability segments. In contrast, the moderate- and low-quality clusters exhibit more heterogeneous label distributions, reflecting gradual transitions and increased variability in signal reliability. The selected thresholds were chosen to reflect established physiological constraints and standard HRV preprocessing practices \cite{peltola2012rr,electrophysiology1996heart}, ensuring robustness to non-physiological artifacts while preserving meaningful cardiac variability.

\begin{figure}[t]
    \centering
    \begin{subfigure}{\linewidth}
        \centering
        \includegraphics[width=\linewidth]{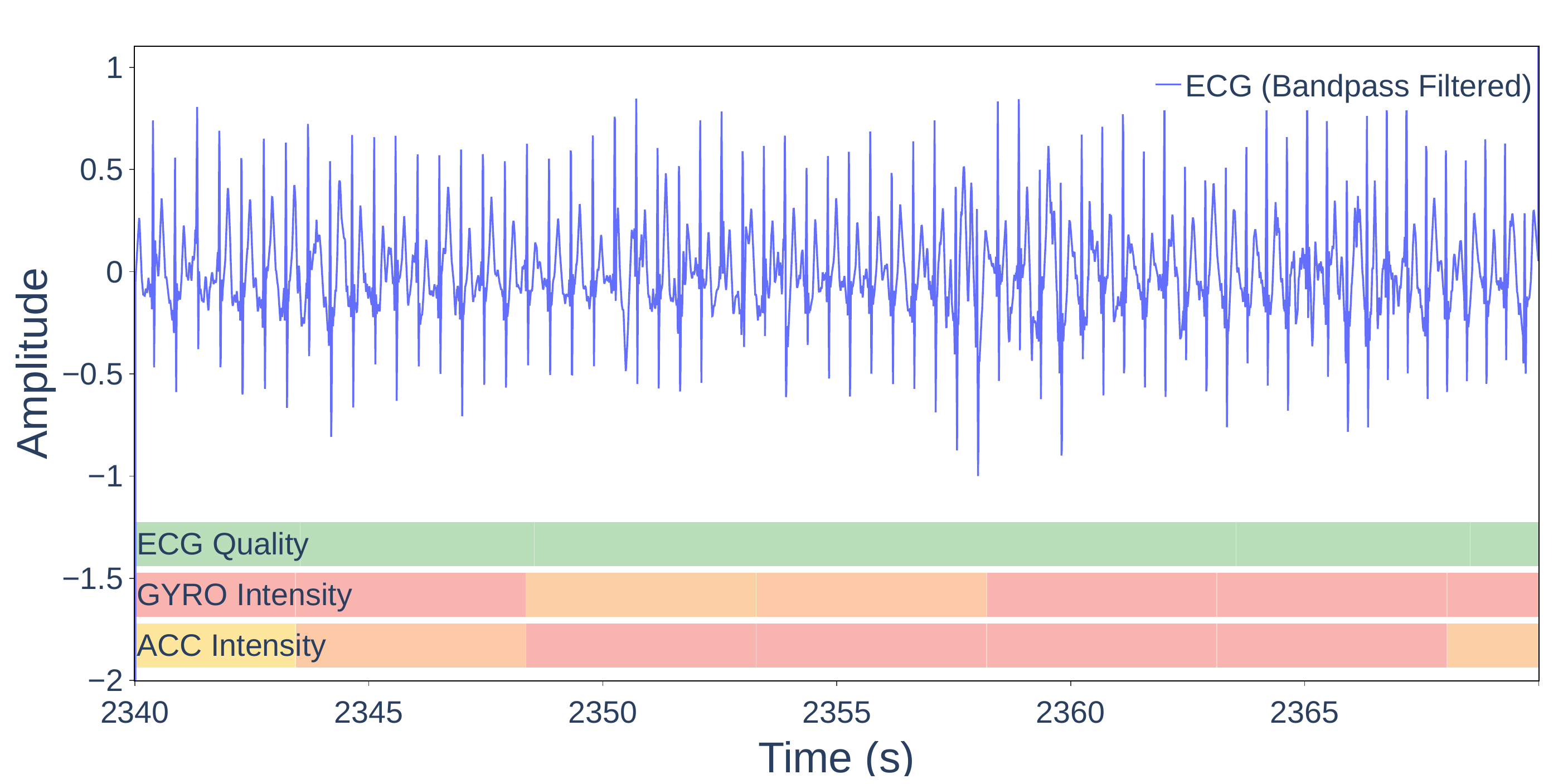}
        \caption{High-intensity segment with high-quality ECG}
    \end{subfigure}

    \vspace{4pt}

    \begin{subfigure}{\linewidth}
        \centering
        \includegraphics[width=\linewidth]{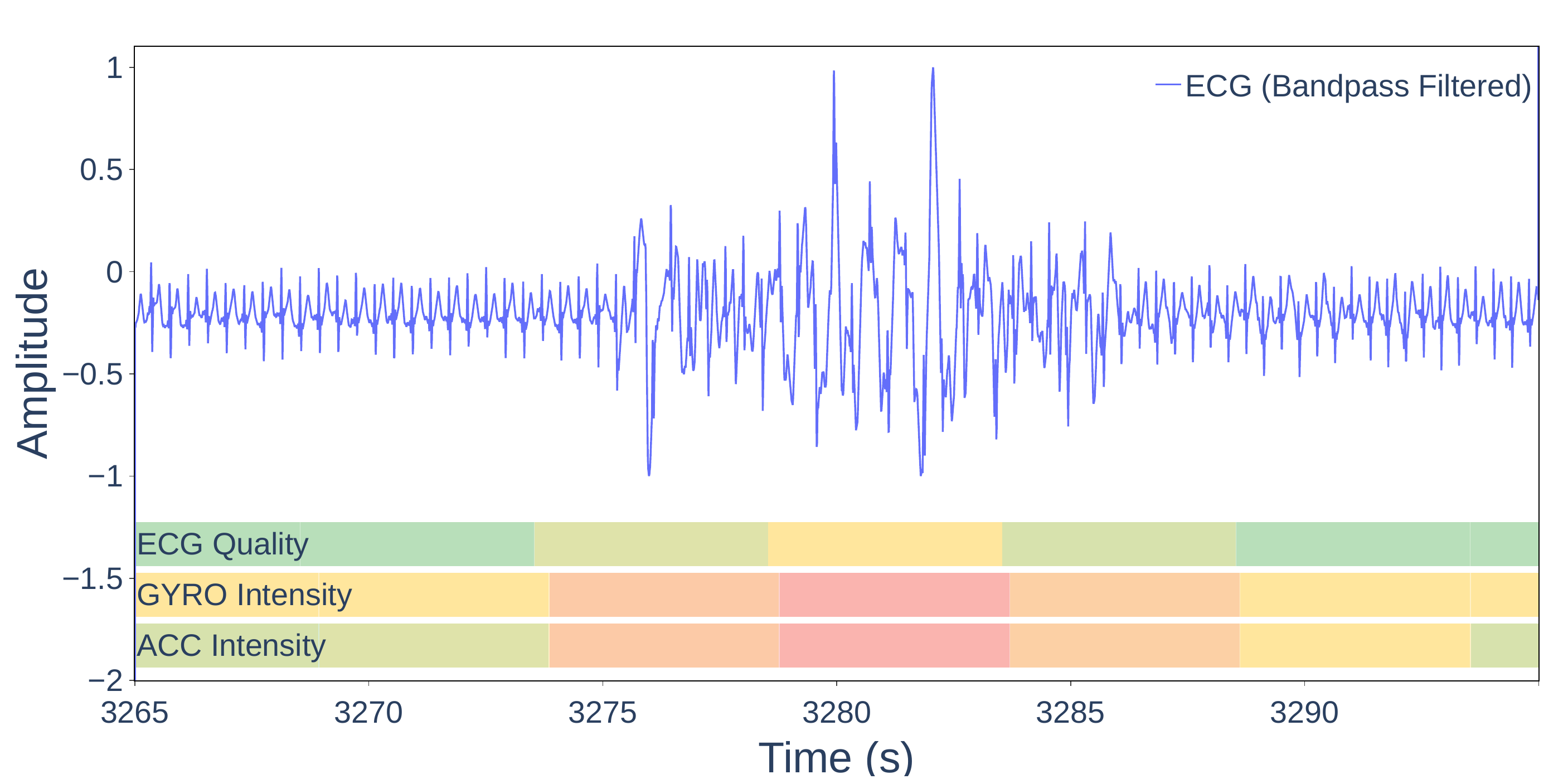}
        \caption{High-intensity segment with moderate-quality ECG}
    \end{subfigure}

    \caption{Representative ECG segments during high-intensity activity. Both segments correspond to elevated motion intensity (ACC and GYRO), but differ in ECG signal quality. The top panel shows a high-quality ECG segment with stable morphology, while the bottom panel shows a moderate-quality segment with increased variability. Colored bands indicate ECG quality (green: high, yellow: moderate) and IMU-derived activity intensity (green: low, yellow: moderate, red: high).}
    \label{fig:ecg_quality_examples}
\end{figure}

The results confirm that clustering-based grouping aligns with physiological signal quality. As shown in Table~\ref{tab:ecg_quality_main}, the high-quality cluster achieves near-perfect RR validity and strong waveform consistency, whereas lower-quality clusters exhibit substantially degraded metrics. This demonstrates that the proposed approach effectively identifies reliable ECG segments without requiring manual annotations, enabling robust HRV analysis in real-world settings. Figure~\ref{fig:ecg_quality_examples} provides representative examples of ECG segments across different quality levels alongside the corresponding activity intensity measurements.

\begin{table}[t]
\centering
\caption{Annotation-free ECG quality metrics used for cluster validation.}
\renewcommand{\arraystretch}{1.3}
\setlength{\tabcolsep}{5pt}

\begin{tabular}{p{3cm} p{5cm}}
\hline
\textbf{Metric} & \textbf{Description} \\
\hline
\strut Valid RR ratio & Fraction of RR intervals within 300--2000ms \\
\strut RR instability ratio & Fraction of abrupt changes between consecutive RR intervals \\
\strut RR CV & Coefficient of variation of RR intervals within each window \\
\strut Beat-template correlation & Average correlation between individual beats and a local template \\
\hline
\end{tabular}

\label{tab:ecg_quality_metrics}
\end{table}

\begin{table}[t]
\centering
\caption{Annotation-free ECG quality metrics across clusters (mean values).}

\setlength{\tabcolsep}{3pt} % default is ~6pt → reduce spacing

\begin{tabular}{lcccc}
\hline
\textbf{Cluster Quality} & \textbf{Valid RR} & \textbf{RR CV} & \textbf{Beat Corr} & \textbf{Windows (\%)} \\
\hline
High     & 0.99 & 0.02 & 0.94 & 63.8 \\
Moderate & 0.40 & 0.25 & 0.82 & 35.6 \\
Low      & 0.88 & 0.33 & 0.61 & 0.6 \\
\hline
\end{tabular}

\label{tab:ecg_quality_main}
\end{table}

\begin{table}[t]
\centering
\caption{Distribution of ECG signal reliability labels across clusters. Values are reported as percentages.}
\setlength{\tabcolsep}{3pt}
\begin{tabular}{lccc}
\hline
\textbf{Cluster Quality} & \textbf{High Reliability} & \textbf{Moderate} & \textbf{Low Reliability} \\
\hline
High     & 96.1 & 3.1 & 0.8 \\
Moderate & 20.4 & 43.1 & 36.5 \\
Low      & 19.8 & 39.1 & 41.1 \\
\hline
\end{tabular}
\label{tab:ecg_quality_labels}
\end{table}

\subsection{Clustering-Based Activity Intensity Segmentation}

To characterize activity intensity from wearable sensor data, a clustering-based segmentation approach was applied to accelerometer (ACC) and gyroscope (GYRO) signals. Data were segmented into overlapping windows of 10~s with a 5~s step, enabling short-term characterization of motion patterns in free-living conditions.

\begin{figure}[t]
    \centering

    \begin{subfigure}{0.49\linewidth}
        \centering
        \includegraphics[width=\linewidth]{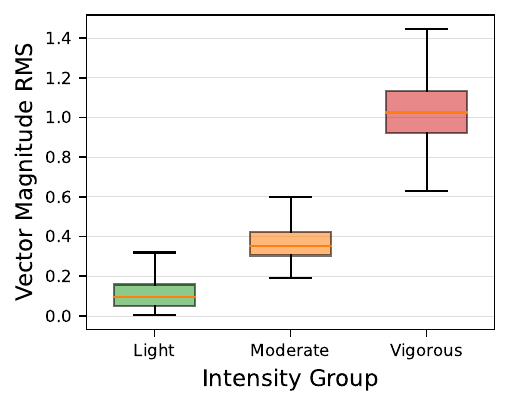}
        \caption{ACC}
    \end{subfigure}
    \hfill
    \begin{subfigure}{0.49\linewidth}
        \centering
        \includegraphics[width=\linewidth]{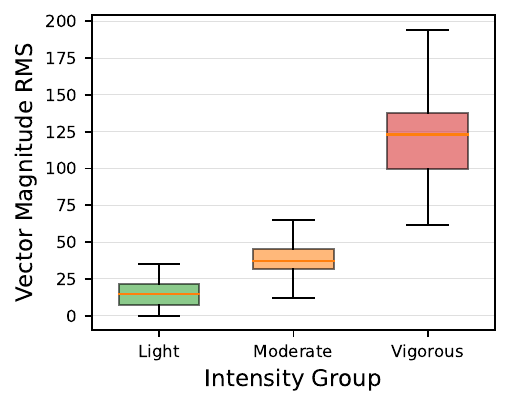}
        \caption{GYRO}
    \end{subfigure}

    \caption{Distribution of vector magnitude RMS (VM RMS) across clustering-derived activity intensity groups. The monotonic increase from light to vigorous activity demonstrates that clustering-based segmentation captures meaningful differences in movement intensity.}

    \label{fig:vm_rms_boxplots}
\end{figure}

For each modality, window-level features were extracted to capture signal properties relevant to activity intensity (see Table~\ref{tab:features}). These features characterize signal magnitude, variability, and motion dynamics, including vector magnitude and Euclidean Norm Minus One (ENMO), which are widely used indicators of physical activity intensity~\cite{marin2020using}. All features were first standardized using z-score normalization to ensure comparability across dimensions. Global K-means clustering with \(K=3\) was then applied to ACC and GYRO features to partition windows into three groups \cite{macqueen1967multivariate}. The resulting clusters were ordered based on signal-energy measures and mapped to \textit{light}, \textit{moderate}, and \textit{vigorous}. The interpretability of this mapping is illustrated in Figure~\ref{fig:vm_rms_boxplots}, which shows the distribution of vector magnitude RMS across the derived intensity groups, demonstrating a clear monotonic increase from light to vigorous activity.

\begin{figure}[t]
    \centering

    \begin{subfigure}{0.49\linewidth}
        \centering
        \includegraphics[width=\linewidth]{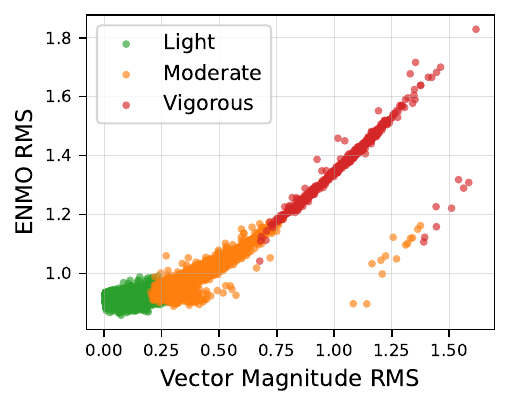}
        \caption{ACC}
    \end{subfigure}
    \hfill
    \begin{subfigure}{0.49\linewidth}
        \centering
        \includegraphics[width=\linewidth]{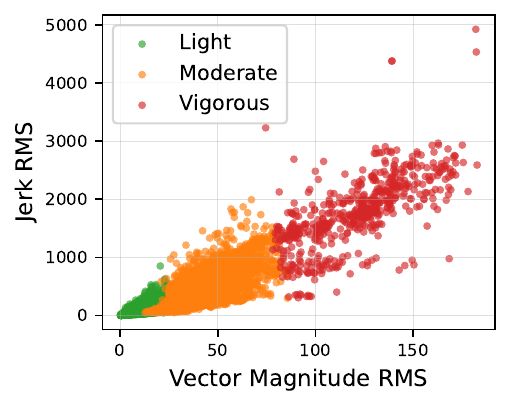}
        \caption{GYRO}
    \end{subfigure}

    \caption{Feature-space representation of clustering-derived activity intensity groups for accelerometer (ACC) and gyroscope (GYRO) signals. Each point represents a windowed segment, colored by cluster assignment. Clear separation across clusters indicates effective discrimination of activity intensity levels based on motion-derived features.}
    
    \label{fig:clustering}
\end{figure}

Figure~\ref{fig:clustering} further illustrates the modality-specific structure of the derived groups, where ACC and GYRO feature spaces exhibit clear separation across activity levels. Clustering performance was evaluated using the Silhouette score~\cite{rousseeuw1987silhouettes} and Davies--Bouldin index~\cite{davies2009cluster}. As shown in Table~\ref{tab:cluster_validation}, both IMU modalities achieve strong separation, supporting the effectiveness of the proposed segmentation approach. Finally, modality-specific segmentations were combined by assigning each window the higher intensity level derived from ACC and GYRO signals, ensuring that periods of increased movement detected by either sensor were retained in the final labeling and reducing the risk of underestimating activity intensity.

Although clustering was performed using K=3 to capture the full spectrum of activity intensities, the vigorous activity group was excluded from downstream statistical analysis. This decision was made because vigorous activity segments were substantially less frequent and exhibited increased motion-related variability and reduced physiological reliability compared to the light and moderate groups. Consequently, subsequent HR and HRV analyses focused on light and moderate activity intensities, where signal quality and sample representation were more reliable.

\begin{table}[t]
\centering
%\scriptsize
\caption{ACC and GYRO feature groups used for activity intensity classification.}
\begin{tabular}{p{2cm} p{5.5cm}}
\hline
\textbf{Category} & \textbf{Features} \\
\hline

Magnitude 
& VM mean, RMS \\

Energy 
& ENMO mean, ENMO RMS, jerk RMS \\

Variability 
& MAD, axis-wise STD \\

Frequency 
& Step-band ratio, tilt variance \\

\hline
\end{tabular}

\vspace{2pt}
\raggedright
\footnotesize{\textit{Abbreviations:} VM (vector magnitude); ENMO (Euclidean norm minus one); MAD (mean absolute deviation); BP (bandpass); HF (high frequency); STD (standard deviation).}

\label{tab:features}
\end{table}

\begin{table}[t]
\centering
\caption{Activity intensity detection using motion sensors with ECG-based physiological validation.}
\begin{tabular*}{\columnwidth}{@{\extracolsep{\fill}}lcc}
\hline
\textbf{Sensor} & \textbf{Silhouette Score} & \textbf{Davies--Bouldin Score} \\
  %     & Score      & Index           \\
\hline
ACC  & 0.43 & 0.80 \\
GYRO & 0.31 & 1.02 \\
\hline
\end{tabular*}
\label{tab:cluster_validation}
\end{table}

\subsection{Peak Detection}
\label{sec:feature_extraction}

Following preprocessing and signal quality validation, physiological features were extracted from ECG signals using a sliding window framework. R-peaks were detected within 10\,s ECG windows using an extended version of our previously proposed framework~\cite{FARROKHI2025109478}. The method enhances the QRS complex through bandpass filtering and wavelet-based decomposition, enabling reliable identification of candidate peaks even in the presence of noise. Peak candidates are then identified using adaptive amplitude thresholding, where detection thresholds are dynamically adjusted based on segment-level signal characteristics.

To ensure robustness under motion artifacts, the initial R-peak detections were evaluated based on temporal consistency, where windows exhibiting implausible or highly irregular RR intervals were identified as unreliable. In such cases, an adaptive fallback mechanism~\cite{de2022adaptive} was triggered to re-estimate R-peaks using relaxed detection thresholds and temporal constraints, enabling recovery from missed or spurious detections. Following re-estimation, peak locations were refined by aligning each detected peak to the nearest local maximum within a small neighborhood to improve temporal precision.

\begin{figure}[t]
    \centering
    \includegraphics[width=\linewidth]{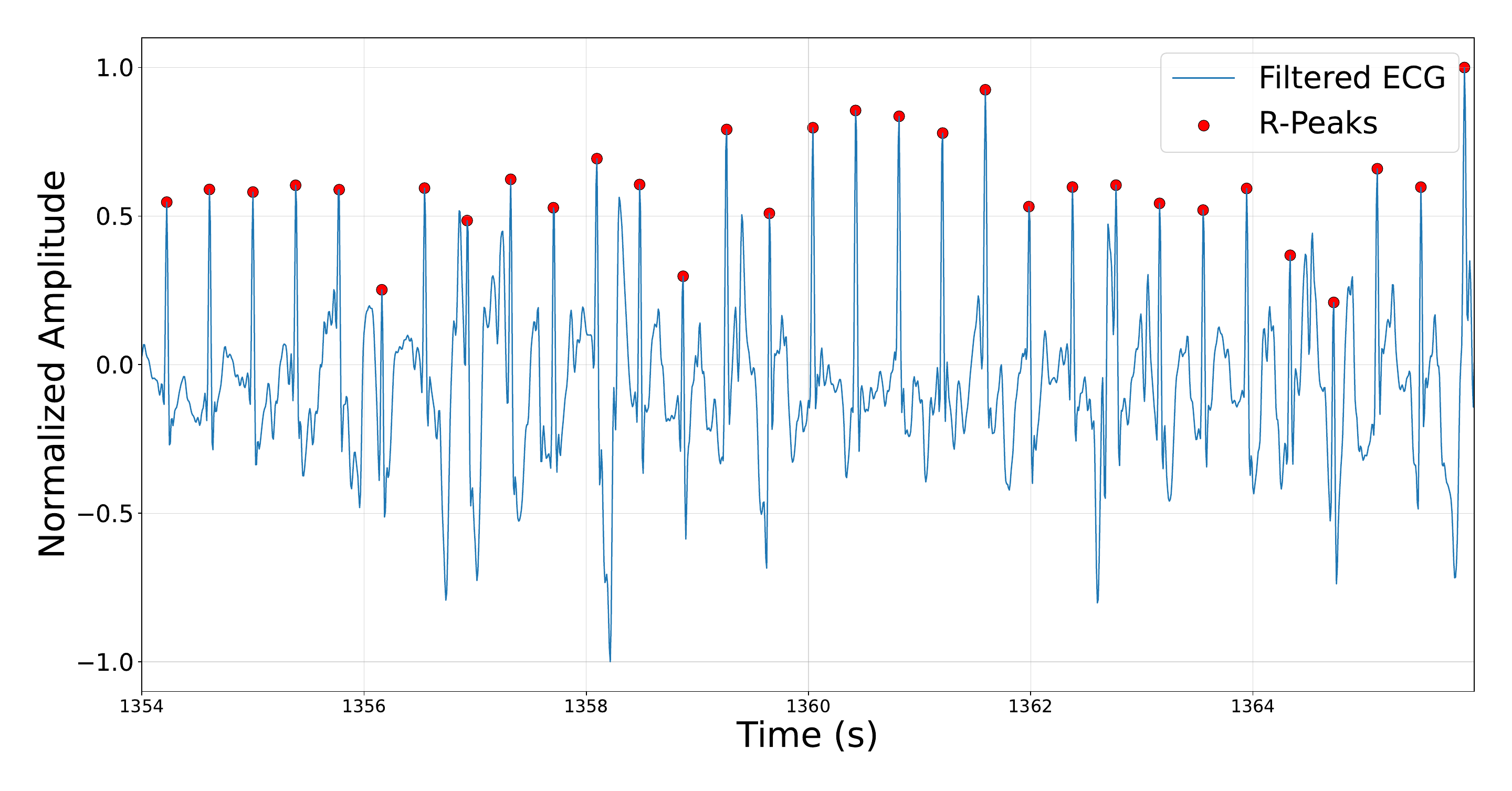}
    \caption{ECG signal with detected R peaks}
    \label{fig:detected_peaks}
\end{figure}

RR intervals were computed from the resulting R-peak sequence and subsequently aggregated over analysis windows of 30 s, 1 min, and 2 min. A final temporal plausibility check was then applied to the detected peaks, ensuring that RR intervals remained within physiologically valid ranges and did not exhibit abrupt, non-physiological beat-to-beat variations. Segments failing prior signal quality criteria or violating these constraints were excluded from further analysis. Figure~\ref{fig:detected_peaks} illustrates an example of the detected R-peaks on a representative ECG segment.

\subsection{Statistical Analysis}

To investigate physiological differences between cancer survivors and healthy controls, heart rate and heart rate variability metrics were analyzed across light and moderate activity intensities. HRV was quantified using the root mean square of successive differences (RMSSD) and the standard deviation of normal-to-normal intervals (SDNN), two widely established time-domain measures of autonomic regulation \cite{shaffer2017overview}. These metrics were computed using sliding windows of 30\,s, 1\,min, and 2\,min to capture cardiovascular responses across multiple temporal scales. Descriptive statistics for all metrics are summarized in Table~\ref{tab:hrv_means}.

To ensure statistically robust and valid comparisons, analyses were conducted at both the participant and window levels. At the participant level, HR and HRV measurements were aggregated within each individual to preserve independence between observations and avoid bias from repeated measures. Group differences were then evaluated using Welch’s t-test, which is appropriate under unequal variances and sample sizes \cite{welch1947generalization}. The resulting test statistics, $p$-values, and effect sizes (Cohen’s $d$) are reported in Table~\ref{tab:participant_stats}.

To further capture temporal dynamics, a complementary window-level analysis was performed in which each valid window was treated as an observation. Given the potential for non-normal distributions, heteroscedasticity, and imbalanced sample sizes at this level, both parametric and nonparametric tests were applied. Specifically, Welch’s t-test was used to assess differences in group means, while the Mann--Whitney U test was used to evaluate distributional differences without assuming normality \cite{mann1947test}. Results of this analysis are presented in Table~\ref{tab:window_level_stats}.

Statistical significance was defined as $p < 0.05$. In addition to hypothesis testing, effect sizes were quantified using Cohen’s $d$ to assess the magnitude of group differences independent of sample size \cite{cohen2013statistical}. This is particularly important in window-level analyses, where large numbers of observations may lead to statistically significant results with limited practical relevance. Accordingly, both statistical significance and effect size are considered when interpreting the results.

Overall, this multi-level statistical framework enables a comprehensive and robust evaluation of differences in HR and HRV between cancer survivors and healthy controls across varying activity intensities and temporal resolutions, while accounting for common challenges in wearable physiological data, including non-normality, unequal variances, and sample size imbalance.

\section{Results and discussion}

This section discusses the differences in cardiac response to physiological exertion between cancer survivors and healthy controls, both during phases of steady-state activity and during activity transitions. First, we analyze heart rate and heart rate variability as a function of activity intensity and window length. Subsequently, we assess statistical significance through comparisons at both the participant and window levels.

\subsection{Heart Rate and HRV Across Activity Intensities}

\begin{figure*}[t]
\centering

\includegraphics[width=0.48\textwidth]{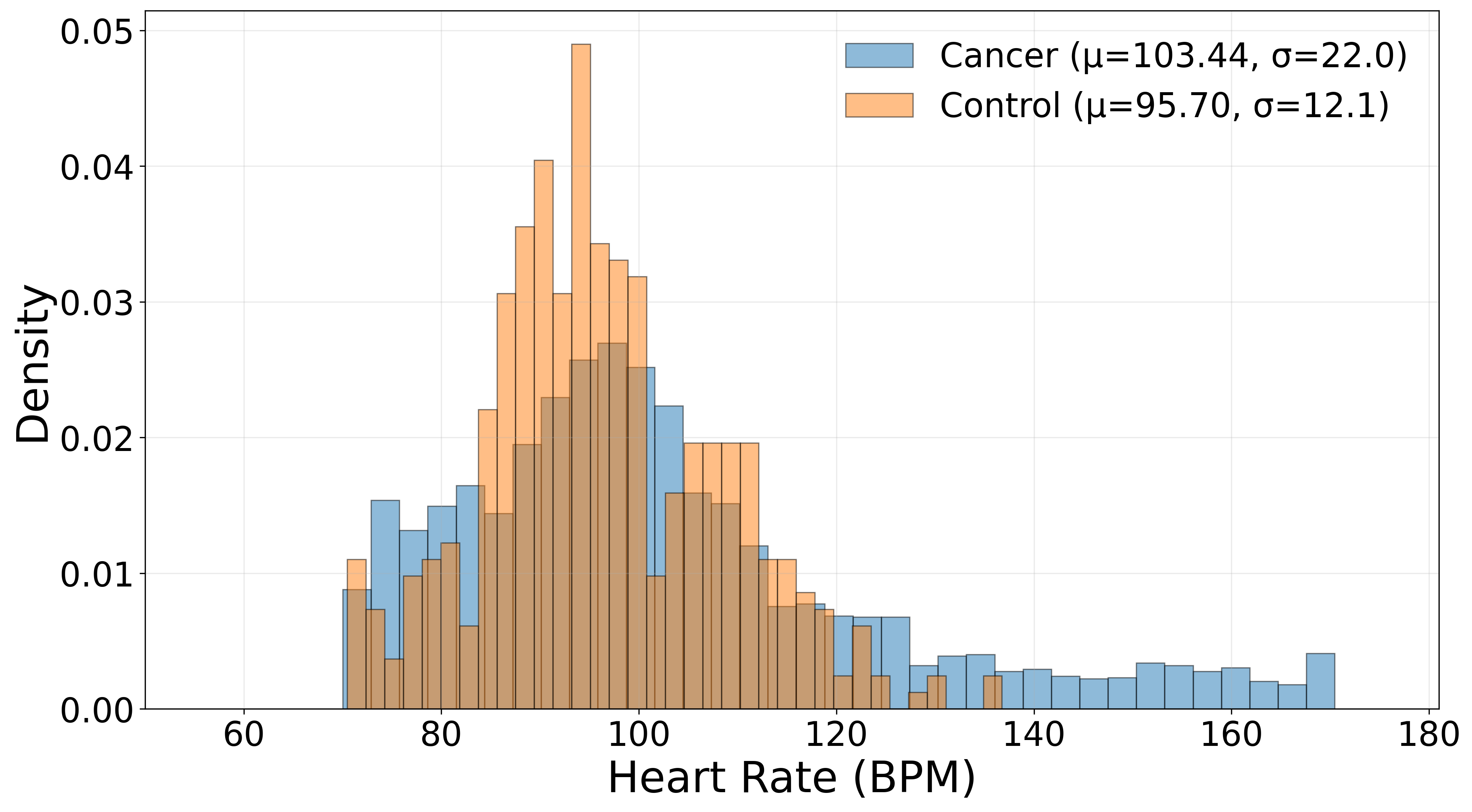}
\hfill
\includegraphics[width=0.48\textwidth]{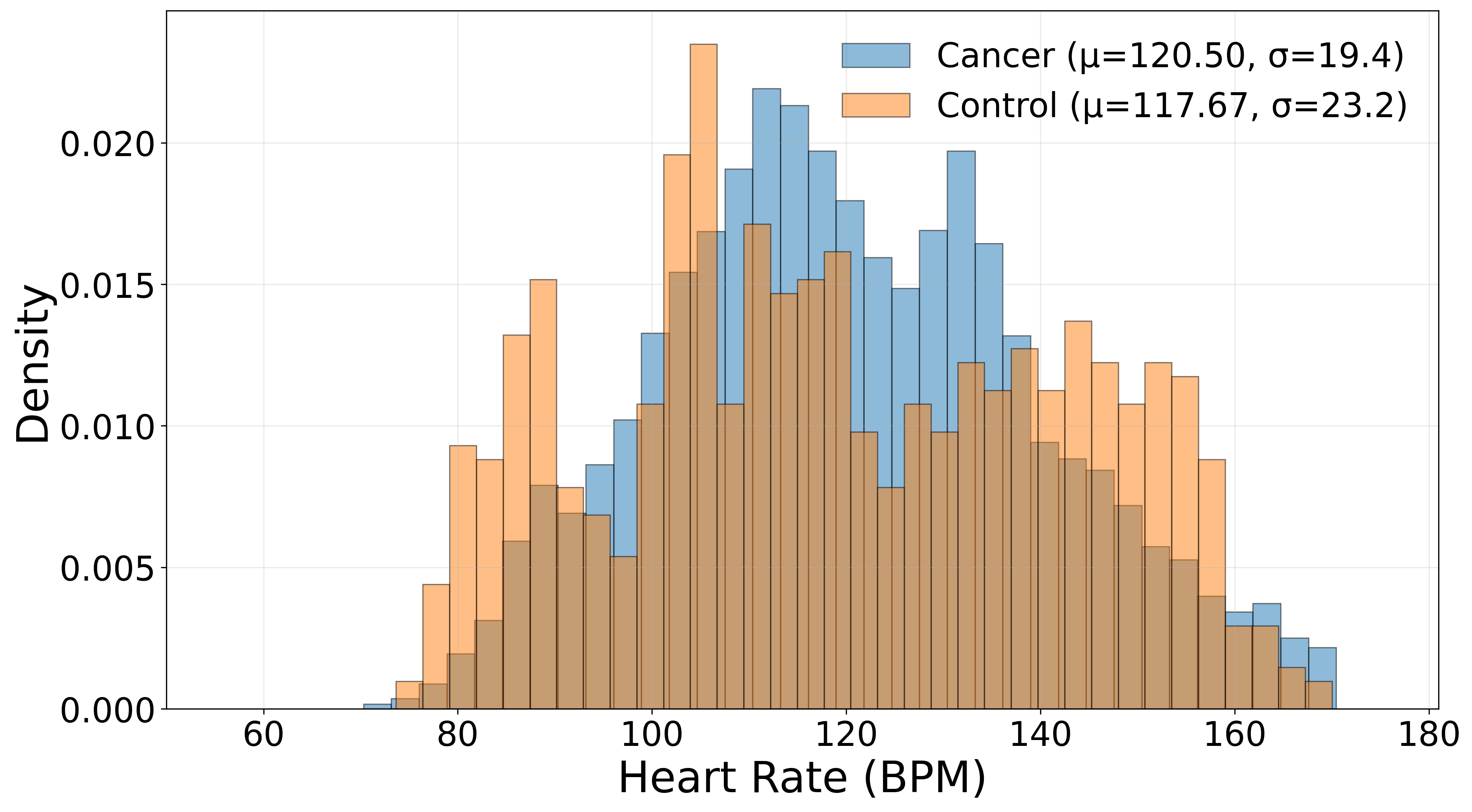}

\caption{
Heart rate distributions for cancer survivors and healthy controls across light and moderate activity intensities. Cancer survivors exhibit elevated heart rates and broader variability across activity levels. Window-level statistical comparisons showed significant differences during light activity ($p \approx 1 \times 10^{-22}$) and moderate activity ($p \approx 0.0017$).
}

\label{fig:hr_distributions}

\end{figure*}

Table~\ref{tab:hrv_means} presents the mean HR and HRV values for both groups across activity intensities. Baseline measurements were available only for the 1\,min analysis windows and are included for reference. As expected, HR increased with activity intensity in both groups, reflecting normal cardiovascular responses to increased physical demand. However, cancer survivors consistently exhibited higher HR values during light activity compared to controls. For example, in the 30\,s windows, cancer survivors showed an average HR of 103.44\,bpm compared to 95.70\,bpm for controls. Figure~\ref{fig:hr_distributions} illustrates the corresponding HR distributions for the 30\,s windows across activity intensities for both groups.

In contrast, HRV metrics were consistently lower in cancer survivors across most activity intensities and window lengths. These differences were particularly pronounced during moderate activity. For instance, in the 1\,min windows during moderate activity, RMSSD averaged 22.09\,ms for cancer survivors compared to 39.68\,ms for controls, while SDNN was 25.14\,ms for cancer survivors and 42.62\,ms for controls. Similar patterns were observed across other window lengths, indicating reduced autonomic variability in cancer survivors during physical activity.

\begin{table*}[t]
\centering
\caption{Comparison of heart rate and HRV metrics between cancer survivors and healthy controls across activity intensities and window lengths. Baseline values are reported only for the 1-minute window. HR is reported in beats per minute (bpm), while RMSSD and SDNN are reported in milliseconds (ms).}
\resizebox{\textwidth}{!}{%
\begin{tabular}{|c|c|ccc|ccc|ccc|}
\hline
\multirow{2}{*}{\textbf{Window}} & \multirow{2}{*}{\textbf{Group}} &
\multicolumn{3}{c|}{\textbf{HR Mean}} &
\multicolumn{3}{c|}{\textbf{RMSSD Mean}} &
\multicolumn{3}{c|}{\textbf{SDNN Mean}} \\
\cline{3-11}
& & Baseline & Light & Moderate
& Baseline & Light & Moderate
& Baseline & Light & Moderate \\
\hline

\multirow{2}{*}{30 s}
& Cancer
& -- & 103.44 & 120.50
& -- & 27.90 & 26.18
& -- & 30.08 & 25.37 \\
& Control
& -- & 95.70 & 117.67
& -- & 35.67 & 56.56
& -- & 44.32 & 50.09 \\
\hline

\multirow{2}{*}{1 min}
& Cancer
& 79.90 & 100.26 & 119.77
& 44.50 & 25.32 & 22.09
& 52.71 & 30.66 & 25.14 \\
& Control
& 83.03 & 93.82 & 116.14
& 43.47 & 31.58 & 39.68
& 58.50 & 44.01 & 42.62 \\
\hline

\multirow{2}{*}{2 min}
& Cancer
& -- & 98.00 & 118.86
& -- & 28.85 & 25.57
& -- & 35.69 & 30.67 \\
& Control
& -- & 91.89 & 114.20
& -- & 31.69 & 45.34
& -- & 47.34 & 48.89 \\
\hline

\end{tabular}
}
\label{tab:hrv_means}
\end{table*}

\subsection{Participant-Level Statistical Comparisons}
To assess participant-level differences between cancer survivors and healthy controls, Welch’s t-tests were conducted, as summarized in Table~\ref{tab:participant_stats}. The most consistent group differences were observed during moderate activity.

For RMSSD, moderate activity yielded statistically significant differences across all window lengths (30\,s: $p=0.04$, $d=-0.63$; 1\,min: $p=0.02$, $d=-0.75$; 2\,min: $p=0.019$, $d=-0.67$), indicating moderate reductions in heart rate variability among cancer survivors. A similar pattern was observed for SDNN, with significant differences during moderate activity (30\,s: $p=0.02$, $d=-0.73$; 1\,min: $p=0.004$, $d=-1.03$; 2\,min: $p=0.003$, $d=-0.98$), where effect sizes were generally larger for SDNN than RMSSD, particularly during light and moderate activity. In contrast, differences during light activity were not statistically significant for either metric across window lengths, with small effect sizes ($|d| \leq 0.45$), indicating minimal group differences under low physical demand. These findings are further examined at a finer temporal resolution using window-level analysis.

\begin{table}[t]
\centering
\caption{Participant-level statistical comparison of HRV metrics between cancer survivors and healthy controls across activity intensities and window lengths using Welch's t-test. Statistically significant results ($p<0.05$) are shown in bold.}
\begin{tabular}{|c|c|c|c|c|c|}
\hline
\textbf{Window} & \textbf{Metric} & \textbf{Intensity} & \textbf{Welch $t$} & \textbf{$p$-value} & \textbf{Cohen's $d$} \\
\hline

\multirow{4}{*}{30 s}
& \multirow{2}{*}{RMSSD}
& Light & -0.41 & 0.68 & -0.11 \\
& & Moderate & -2.11 & \textbf{0.04} & -0.63 \\
\cline{2-6}
& \multirow{2}{*}{SDNN}
& Light & -0.92 & 0.36 & -0.33 \\
& & Moderate & -2.43 & \textbf{0.02} & -0.73 \\
\hline

\multirow{4}{*}{1 min}
& \multirow{2}{*}{RMSSD}
& Light & -1.04 & 0.31 & -0.30 \\
& & Moderate & -2.43 & \textbf{0.02} & -0.75 \\
\cline{2-6}
& \multirow{2}{*}{SDNN}
& Light & -1.09 & 0.28 & -0.32 \\
& & Moderate & -3.15 & \textbf{0.004} & -1.03 \\
\hline

\multirow{4}{*}{2 min}
& \multirow{2}{*}{RMSSD}
& Light & -0.88 & 0.40 & -0.36 \\
& & Moderate & -2.41 & \textbf{0.019} & -0.67 \\
\cline{2-6}
& \multirow{2}{*}{SDNN}
& Light & -1.06 & 0.31 & -0.45 \\
& & Moderate & -3.16 & \textbf{0.003} & -0.98 \\
\hline

\end{tabular}

\label{tab:participant_stats}
\end{table}
\subsection{Window-Level Statistical Comparisons}

Window-level analysis further characterizes group differences across activity intensities. Across window sizes, HRV metrics were generally lower in cancer survivors compared to healthy controls, with the magnitude of differences varying by activity intensity.

Both Welch’s t-test and the Mann--Whitney U test indicated statistically significant differences between groups for nearly all comparisons ($p < 0.05$). However, due to the large number of window-level observations and repeated measurements within participants, statistical significance is expected and may be inflated. Therefore, interpretation focuses primarily on effect sizes, which provide a more meaningful measure of group differences.

\begin{figure}[t]
    \centering
    \begin{subfigure}{0.48\linewidth}
        \includegraphics[width=\linewidth]{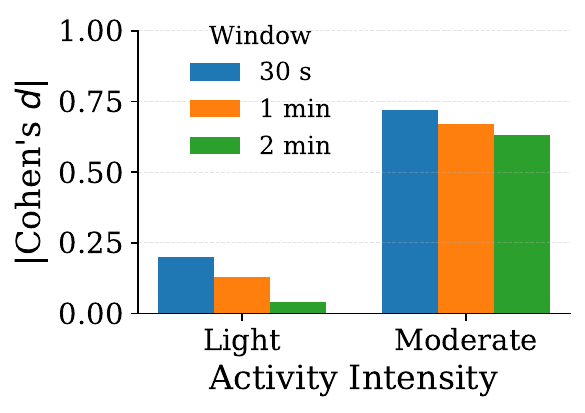}
        \caption{RMSSD}
    \end{subfigure}
    \hfill
    \begin{subfigure}{0.48\linewidth}
        \includegraphics[width=\linewidth]{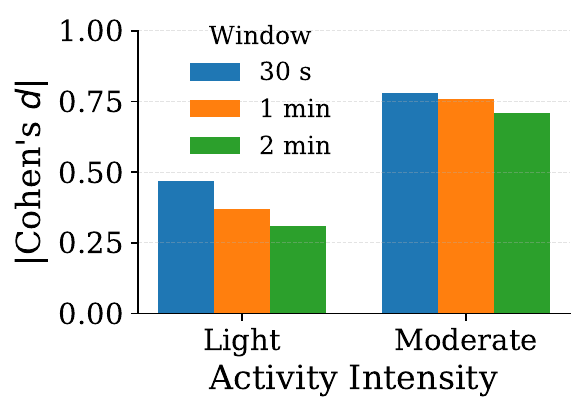}
        \caption{SDNN}
    \end{subfigure}
    \caption{Magnitude of effect sizes ($|$Cohen’s $d|$) across light and moderate activity intensities and window lengths. Moderate activity consistently shows stronger group differences between cancer survivors and controls, while light activity shows weaker separation.}
    \label{fig:effect_sizes}
\end{figure}

As illustrated in Figure~\ref{fig:effect_sizes}, effect size analysis highlights consistent activity-dependent differences between groups. During light activity, effect sizes are negligible to small (e.g., RMSSD $d=-0.20$ for 30\,s windows, $d=-0.13$ for 1\,min windows, and $d=-0.04$ for 2\,min windows), indicating weak separation between groups under low physical demand. In contrast, moderate activity produces consistently moderate-to-large effect sizes (e.g., RMSSD $d=-0.72$, $d=-0.67$, and $d=-0.63$ for 30\,s, 1\,min, and 2\,min windows, respectively), demonstrating a pronounced reduction in HRV among cancer survivors. SDNN follows the same trend, with small differences during light activity (e.g., $d=-0.47$, $d=-0.37$, and $d=-0.31$) and consistently larger effects during moderate activity (e.g., $d=-0.78$, $d=-0.76$, and $d=-0.71$). Consistent with the participant-level analysis, SDNN generally exhibits larger effect sizes than RMSSD during light and moderate activity. Detailed window-level statistics, including the number of observations per group, are reported in Table~\ref{tab:window_level_stats} for completeness.

Overall, moderate activity provides the most robust and consistent separation between groups, indicating that autonomic dysfunction in cancer survivors is most detectable under real-world conditions and becomes more apparent under increased physiological demand.

\begin{table}[t]
\centering
\caption{Window-level comparison of HRV metrics between cancer survivors and healthy controls across window lengths and activity intensities. Values are reported as Cohen's $d$ effect sizes.}

\setlength{\tabcolsep}{3.5pt}
\renewcommand{\arraystretch}{1.15}
%\small

%\begin{tabular}{ccccccc}
\begin{tabular}{|c|c|c|cc|c|c|}
\hline
\textbf{Window} & \textbf{Metric} & \textbf{Intensity} &\textbf{ Cancer $n$} & \textbf{Control $n$} & \textbf{$d$} & \textbf{ES} \\
\hline

\multirow{4}{*}{30 s}
& \multirow{2}{*}{RMSSD}
& Light & 4147 & 458 & -0.20 & S \\
& & Moderate & 11063 & 869 & -0.72 & M--L \\
%& & V & 499 & 62 & -0.98 & L \\
\cline{2-7}
& \multirow{2}{*}{SDNN}
& Light & 4147 & 458 & -0.47 & S--M \\
& & Moderate & 11063 & 869 & -0.78 & L \\
%& & V & 499 & 62 & -0.86 & L \\
\hline

\multirow{4}{*}{1 min}
& \multirow{2}{*}{RMSSD}
& Light & 1479 & 229 & -0.13 & S \\
& & Moderate & 6219 & 660 & -0.67 & M \\
%& & V & 347 & 56 & -1.06 & L \\
\cline{2-7}
& \multirow{2}{*}{SDNN}
& Light & 1479 & 229 & -0.37 & S \\
& & Moderate & 6219 & 660 & -0.76 & L \\
%& & V & 347 & 56 & -0.87 & L \\
\hline

\multirow{4}{*}{2 min}
& \multirow{2}{*}{RMSSD}
& Light & 492 & 167 & -0.04 & N \\
& & Moderate & 3430 & 550 & -0.63 & M \\
%& & V & 248 & 41 & -1.03 & L \\
\cline{2-7}
& \multirow{2}{*}{SDNN}
& Light & 492 & 167 & -0.31 & S \\
& & Moderate & 3430 & 550 & -0.71 & M--L \\
%& & V & 248 & 41 & -0.79 & L \\
\hline

\end{tabular}

\vspace{2pt}
\footnotesize{ES: effect size (N: negligible, S: small, M: moderate, L: large).}

\label{tab:window_level_stats}
\end{table}

\subsection{Discussion and Comparison with Prior Work}

The results demonstrate consistent and statistically supported differences in autonomic regulation between cancer survivors and healthy controls across all window lengths, with the strongest and most consistent findings observed during moderate activity. As shown in Table~\ref{tab:hrv_means}, cancer survivors exhibit lower HRV compared to controls across all conditions. For example, in 1-minute windows during moderate activity, RMSSD decreases from 39.68~ms in controls to 22.09~ms in cancer survivors, while SDNN decreases from 42.62~ms to 25.14~ms. These reductions persist across 30-second and 2-minute windows, indicating that the observed group differences are robust to temporal resolution.

These findings are consistent with prior clinical studies reporting altered autonomic regulation in cancer populations. Sousa et al.~\cite{sousa2024physical} conducted a prospective case-control study with pre- and post-chemotherapy assessments under controlled resting conditions, evaluating HRV using standard time- and frequency-domain metrics. Their results demonstrated significant reductions in HRV following chemotherapy, with RMSSD decreasing from 44.43$\pm$25.27~ms to 14.89$\pm$8.28~ms and SDNN from 72.14$\pm$61.37~ms to 26.30$\pm$10.37~ms in the cancer group, while the control group showed opposite trends, indicating impaired parasympathetic modulation and overall autonomic imbalance.

Similarly, Majerova et al.~\cite{majerova2022increased} evaluated HRV using 5-minute resting recordings in a controlled seated environment, analyzing time-domain, frequency-domain, and nonlinear metrics. While their time-domain analysis did not reveal significant differences in RMSSD and SDNN between groups, nonlinear measures identified increased sympathetic modulation in cancer survivors, reflected by higher 0V\% values (16.17$\pm$6.44\% vs. 26.23$\pm$9.7\%) and reduced signal complexity indicated by lower sample entropy (1.86$\pm$0.21 vs. 1.60--1.62). The absence of significant differences in conventional time-domain metrics suggests that resting HRV analysis may underestimate autonomic alterations in this population. This is consistent with the present findings, where baseline HRV values in the 1-minute windows were relatively comparable between groups (RMSSD: 44.50~ms vs. 43.47~ms; SDNN: 52.71~ms vs. 58.50~ms), while substantially larger differences emerged during moderate activity.

The present results reconcile these findings by demonstrating that differences in RMSSD and SDNN, which may not be detectable under resting conditions, become evident under moderate activity. \textbf{This indicates that activity context plays a significant role in revealing autonomic dysfunction and that physiological differences between groups are context-dependent, emerging more clearly under increased physiological demand.}

Beyond baseline differences, a key observation of this study is the distinct manner in which HRV responds to increasing activity intensity across groups. In healthy controls, short-term variability increases from light to moderate activity, as reflected by an increase in RMSSD from 31.58~ms to 39.68~ms, while SDNN remains relatively stable (44.01~ms to 42.62~ms). In contrast, cancer survivors exhibit a blunted response, with both RMSSD and SDNN decreasing from light to moderate activity (RMSSD: 25.32~ms to 22.09~ms; SDNN: 30.66~ms to 25.14~ms). This behavior contrasts with findings from controlled exercise studies, where HRV typically decreases following high-intensity exertion due to increased sympathetic activation. For example, Wang et al.~\cite{10.3389/fphys.2024.1462082} evaluated HRV using a structured treadmill-based Bruce protocol, with measurements obtained from 5-minute quiet-sitting recordings before exercise and during recovery after exhaustive exertion. Their results showed marked reductions in HRV, with RMSSD decreasing from 39.05~ms to 6.19~ms and SDNN from 40.43~ms to 8.70~ms ($p<0.01$), indicating a shift toward sympathetic dominance. Unlike such controlled protocols with fixed pacing and continuous exertion, the present study was conducted in free-living environments characterized by dynamic and non-stationary activity patterns, which may contribute to greater short-term variability in HRV measurements during moderate activity.

In contrast, the present study analyzes short-duration windows in free-living environments characterized by continuous movement and non-stationary physiological signals. Under these conditions, RMSSD increases in healthy controls from light to moderate activity, which may reflect dynamic cardiovascular adjustments and transient physiological variability. However, cancer survivors fail to exhibit this adaptive response, further supporting the presence of impaired autonomic regulation under real-world conditions.

An additional observation is the influence of window length on HRV estimation. Shorter windows (30\,s) exhibit higher variability and inflated HRV values due to increased sensitivity to transient fluctuations. For example, RMSSD in the control group during moderate activity decreases from 56.56~ms in 30-second windows to 39.68~ms in 1-minute windows and stabilizes near 45.34~ms in 2-minute windows. While longer windows improve stability, they may attenuate short-term dynamics, highlighting a trade-off between temporal resolution and statistical reliability. In this study, 1-minute windows provide a practical balance for capturing dynamic HRV behavior in free-living conditions.

From a physiological perspective, the reduced HRV observed in cancer survivors is consistent with increased sympathetic activation and reduced parasympathetic modulation. These alterations are commonly linked to chronic inflammation, neuroendocrine dysregulation, chemotherapy-induced cardiotoxicity, and surgical stress responses, all of which contribute to long-term autonomic imbalance.

From an application perspective, these findings have important implications for real-world physiological monitoring. The results suggest that reliable assessment of autonomic function in wearable settings requires not only robust signal processing, but also explicit consideration of activity context and signal quality. In particular, moderate activity emerges as a physiologically informative regime for detecting autonomic dysfunction, indicating that continuous monitoring systems may benefit from prioritizing such segments for analysis. This has potential applications in remote patient monitoring, early detection of cardiovascular dysregulation, and personalized health assessment in clinical populations.

Overall, these findings demonstrate that autonomic dysfunction in cancer survivors is both persistent and context-dependent, and highlight the importance of activity-aware HRV analysis for accurate and meaningful physiological monitoring in real-world wearable environments.

\section{Conclusion}

This study presents a unified, quality-aware multimodal framework for analyzing heart rate and heart rate variability in free-living environments using wearable ECG and inertial sensor data. By jointly integrating activity intensity segmentation, ECG signal quality assessment, and physiological feature extraction, the proposed approach enables robust and reliable analysis of cardiovascular dynamics under real-world conditions. The results demonstrate consistent and statistically supported differences in autonomic regulation between cancer survivors and healthy controls. In particular, \textbf{cancer survivors exhibit elevated HR and reduced HRV across activity intensities}. These differences are most pronounced during moderate activity, which emerges as the most informative regime for detecting autonomic dysfunction in real-world conditions. These findings indicate that autonomic dysfunction in this population is not only persistent, but also strongly dependent on activity context, underscoring the importance of activity-aware analysis for meaningful interpretation of wearable physiological data.

While the proposed framework demonstrates strong performance, several limitations should be considered. The dataset is imbalanced in recording duration between groups, with a larger proportion of data available for cancer survivors, which may influence statistical comparisons. In addition, there is a substantial age imbalance between the cancer survivor and control groups, and since HRV is known to decrease with age, part of the observed physiological differences may reflect age-related autonomic changes in addition to cancer survivorship effects. The free-living nature of the data introduces variability in behavior, sensor placement, and environmental conditions, which, while enhancing ecological validity, may affect reproducibility.

Future work will focus on extending the framework to larger and more diverse populations, incorporating additional physiological modalities, and exploring learning-based approaches for joint modeling of signal quality and physiological features. These directions will further support the development of reliable and scalable wearable systems for real-world cardiovascular monitoring and personalized health assessment.

\balance
\bibliographystyle{IEEEtran}
\bibliography{bibliography}
% that's all folks
\end{document}